\documentclass[12pt,preprint]{aastex}
\usepackage{color}
\usepackage{graphicx}
\usepackage{epsfig}
\usepackage{multirow}
\usepackage{footnote}

\newcommand{\kms}{\mbox{km s$^{-1}$}}

\newcommand{\um}{\mbox{$\mu$m}}

\shorttitle{Terrestrial Planet Formation around Sun-like Stars}
\shortauthors{C. Melis et al.}

\begin{document}


\title{The Age of the HD 15407 System and the Epoch of Final Catastrophic Mass Accretion onto Terrestrial Planets around Sun-like Stars}


\author{C. Melis\altaffilmark{1}, B. Zuckerman\altaffilmark{2}, Joseph H. Rhee\altaffilmark{2}, Inseok Song\altaffilmark{3}}
\email{cmelis@ucsd.edu}


\altaffiltext{1}{CASS Postdoctoral Fellow $-$Center for Astrophysics and Space Sciences, University of California, San Diego, CA 92093-0424, USA}
\altaffiltext{2}{Department of Physics and Astronomy, University of California,
Los Angeles, CA 90095-1547, USA}
\altaffiltext{3}{Department of Physics and Astronomy, University of Georgia, Athens, GA 30602-2451, USA}


\begin{abstract}
From optical spectroscopic measurements we determine that 
the HD 15407 binary system is $\sim$80 Myr old. The primary,
HD 15407A (spectral type F5~V), exhibits strong mid-infrared
excess emission indicative of a recent catastrophic collision between 
rocky planetary embryos or planets in its inner planetary system. Synthesis of all known 
stars with large quantities of dust in their terrestrial planet zone indicates that
for stars of roughly Solar mass this warm dust phenomenon
occurs at ages between 30 and 100 Myr. In contrast, for stars of a few
Solar masses, the dominant era of the final assembling of rocky planets
occurs earlier, between 10 and 30 Myr age. The incidence of the warm dust
phenomenon, when compared against models for the formation of rocky
terrestrial-like bodies, implies that rocky planet formation in the terrestrial
planet zone around Sun-like stars is common.
\end{abstract}


\keywords{circumstellar matter --- infrared: planetary systems --- planets and satellites: formation --- stars: individual (HD 15407A, HD15407B) --- stars: kinematics and dynamics}



\section{Introduction}

Investigating the formation of terrestrial planets outside of the Solar System
can give insight into the origin of the Earth-Moon system and whether similar
planetary system architectures form around other stars. The study
of extra-solar terrestrial planet formation must currently be carried out through indirect
searches for the dusty aftermath of formation events. Since the dust is located
in the terrestrial planet zone, it will have temperatures of $\sim$300 K
and will emit most of its thermal radiation at $\sim$10 $\mu$m. Stars actively
undergoing terrestrial planet formation will have substantial amounts of
warm dust orbiting in their terrestrial planet zones and will thus show
mid-infrared emission in excess of what one would expect from
the star alone.
Properly establishing the age of such mid-infrared excess systems is 
necessary for comparison with expectations from planet formation
theories and simulations.

We were alerted to excess mid-infrared emission around the star HD 15407A 
through a paper entitled ``Detection of silica dust in the debris disk around HD 15407''
presented at an October 2009 Spitzer meeting by H. Fujiwara
(E. Becklin, personal communication 2009).
At this meeting, the age for HD 15407A was suggested to be $\sim$2 Gyr. We
felt this age could be in error, and initiated our own follow-up observations
to independently constrain HD 15407A's age. Here we report
these observations, our revised value for the age of the HD 15407 binary system,
and a discussion of terrestrial planet zone collisions around adolescent
Solar mass stars.


\section{Observations}

Both components of the HD 15407 system were observed 
in cloudy conditions at Mauna Kea Observatory with
the Keck~I telescope and HIRES echelle spectrometer \citep{vogt94}. 
With the HIRES red collimator wavelengths between 4680
and 8700 \AA\ were covered at a resolution of 50,000 (as determined 
from the FWHM of single
ThAr arclines). The final signal-to-noise ratio per pixel, measured at 
6700 \AA, is 350 for HD 15407A and 200 for HD 15407B.  Spectral data 
were reduced 
and extracted using both the {\it MAKEE} software package and standard IRAF
tasks and echelle reduction procedures. 




\section{Results}

\subsection{HD 15407 Stellar Parameters}
\label{secsptypes}

Our star of interest, HD 15407A, is part of a visual double star system.
Table \ref{tabpars}
contains information about HD 15407A, the dusty primary star, 
and HD 15407B, the secondary component of the HD 15407
system. For both stars we employ line-depth ratios to roughly 
determine absolute temperature and, for HD 15407B, luminosity. 
Those interested in the details of determining such parameters
are referred to \citet{melis09}.

We deduce that HD 15407A has a temperature similar to that of
an F5-type star; its broad-band colors (BVIJHK$_{\rm s}$)
are in agreement with this temperature class determination. This suggests that
HD 15407A is not significantly reddened by its circumstellar material and we
adopt a temperature class of F5$\pm$1 subclass.
Spectral analysis
suggests that HD 15407B is a dwarf star with a temperature
class of K3$\pm$1 subclass and an effective
temperature of 4600 K. Broad-band colors for HD 15407B 
instead suggest that this dwarf star has an earlier spectral type of K0-K1. 
We adopt a spectral type of K2~V with
an uncertainty of $\pm$1 temperature subclass.

\subsection{Excess Infrared Emission from HD 15407A}
\label{secirex}

\begin{figure}
 \includegraphics[width=160mm]{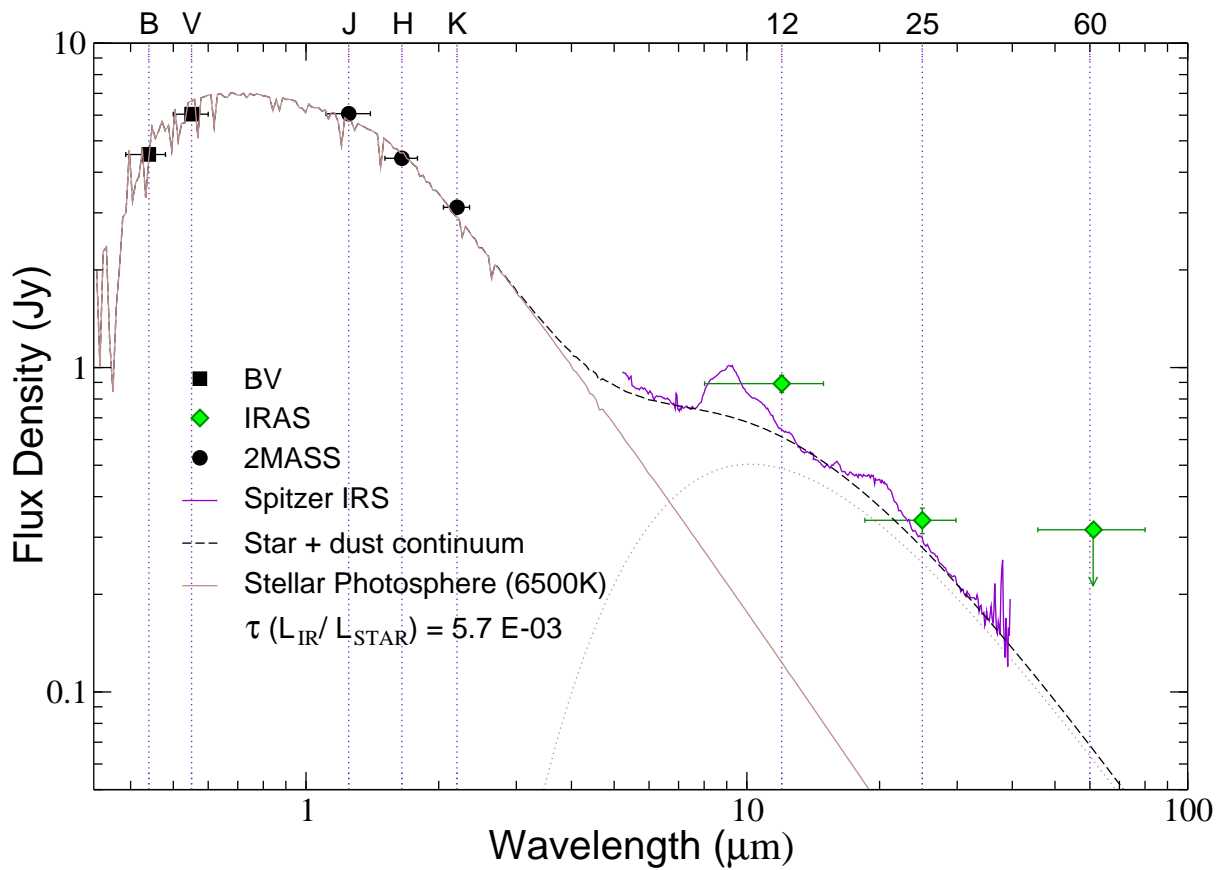}
\caption{\label{figirex}
         Significant mid-infrared excess emission is detected towards HD 15407A. 
         The solid brown curve is a synthetic stellar
         spectrum \citep{hau99} for a 6500 K effective temperature star. 
         The brown dotted curve 
         represents a dust continuum emission blackbody fit to a dust 
         temperature of 500 K. The black dashed line, the
         sum of the above two curves, is a reasonable fit to all data
         points. The total luminosity of the excess, determined
         by integrating under the data points between 1 \um\ and 100 \um , 
         is $\sim$0.6\% of  the luminosity of the star. The BV data points 
         are from the Tycho-2 catalog. The 
         green data points are from the IRAS Faint Source Reject Catalog and have
         not been color corrected. The horizontal
         bars indicate the filter bandwidths. Overplotted is the reduced, extracted,
         calibrated Spitzer IRS spectrum.} 
\end{figure}

\begin{figure}
\centering
\begin{minipage}[t!]{105mm}
 \includegraphics[width=105mm]{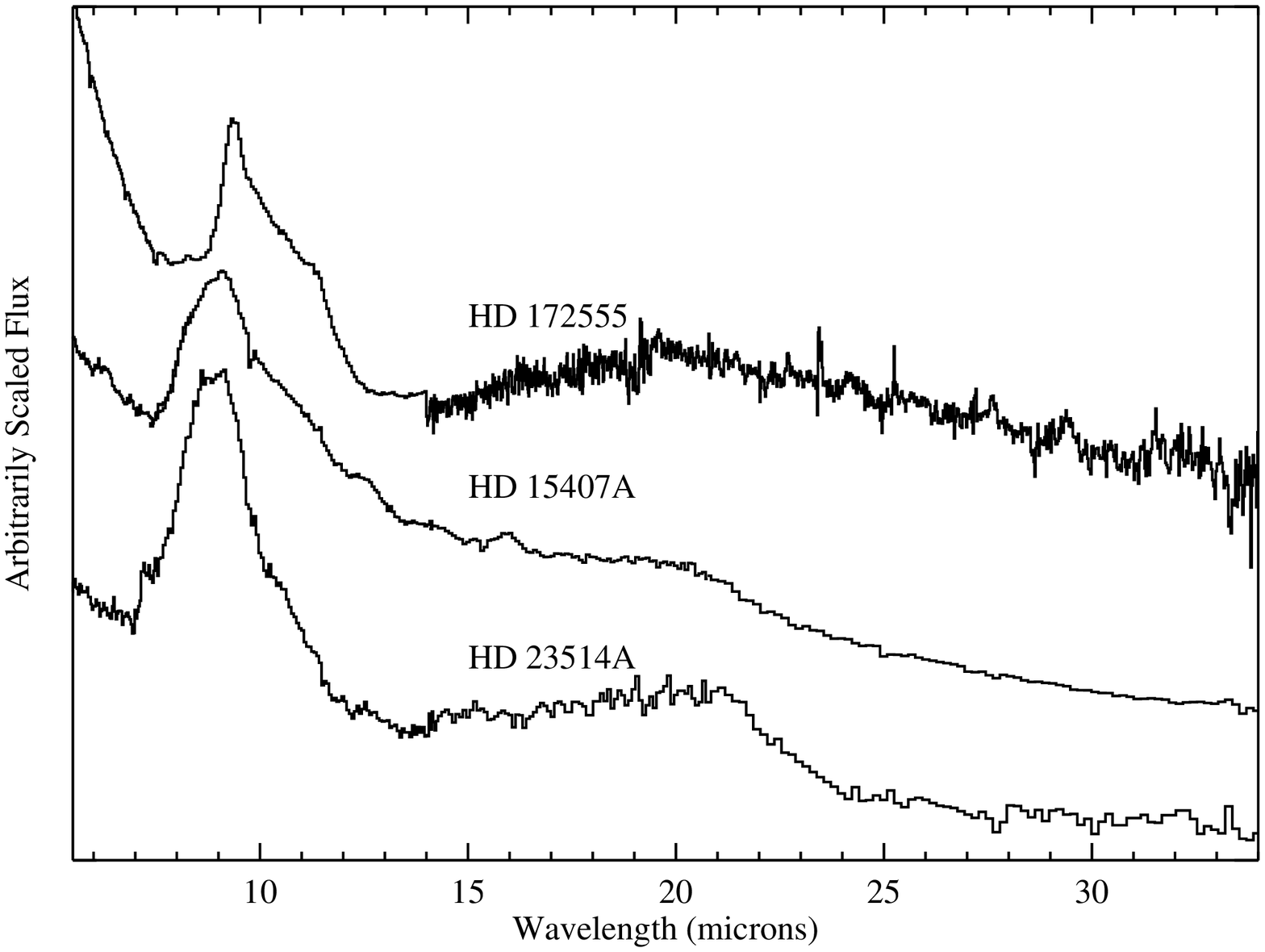}
\end{minipage}
\begin{minipage}[b!]{105mm}
 \includegraphics[width=105mm]{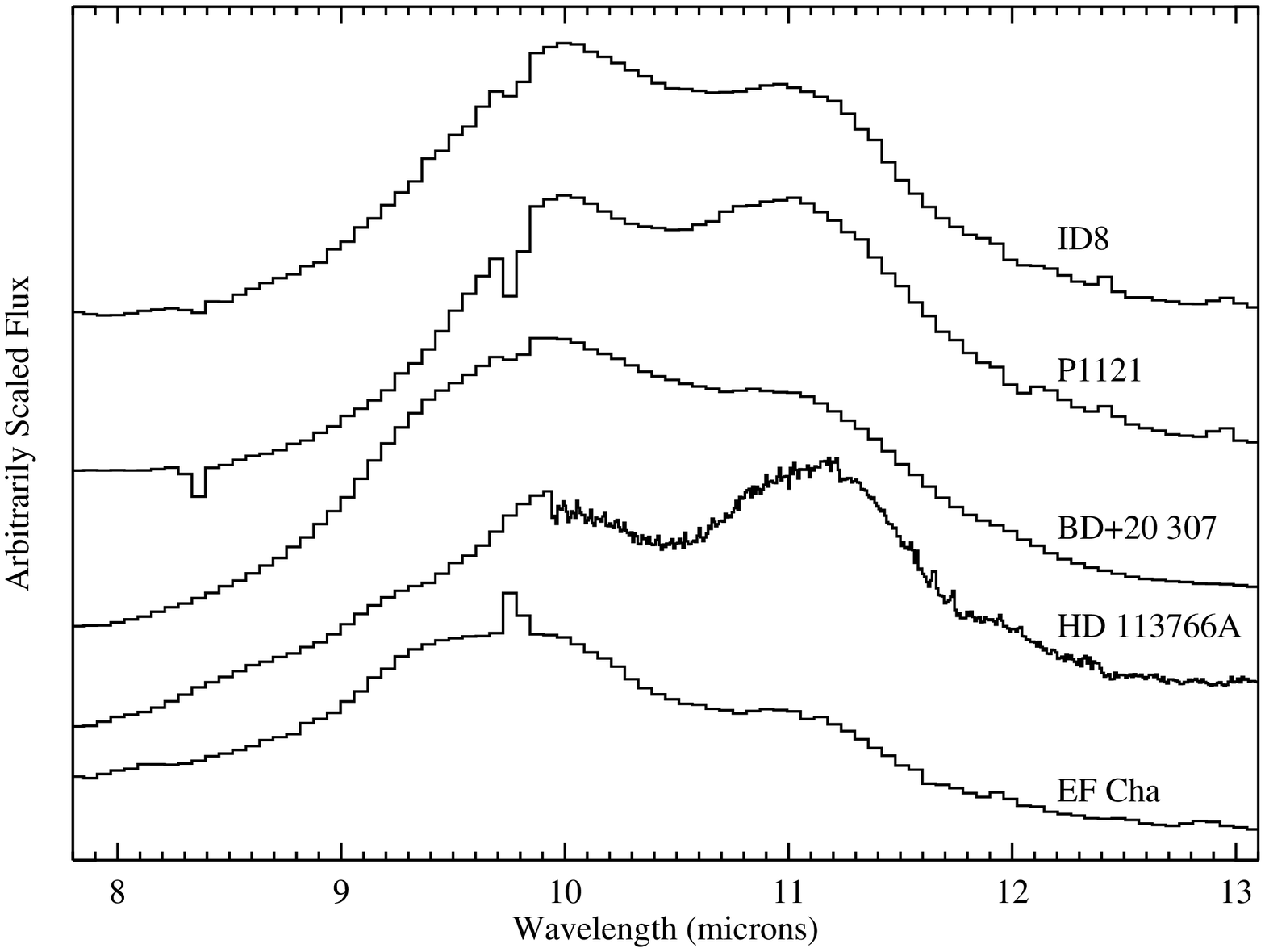}
\end{minipage} 
\caption{\label{figdusty} {\it Top Panel:} Comparison of HD 15407A's IRS spectrum to those of other stars that
               host 9 $\mu$m spectral features. 
               HD 15407A ($\tau$$\sim$0.6\%) and HD 23514A 
               \citep[$\tau$$\sim$2\%; see][D. Rodriguez {\it et al.}\ 2010 submitted]{rhee08} 
               are $\sim$100 Myr old
               mid-F type stars while HD 172555 ($\tau$$\sim$0.08\%) is a $\sim$12 Myr old 
               late-A type star. HD 172555's IRS spectrum has
               been analyzed in great detail and is dominated by features from
               amorphous olivine and pyroxene,
               crystalline pyroxene, and silica \citep{chen06,lisse09}; the latter dust species is
               suggested to result from high velocity impacts. 
               {\it Bottom Panel:} IRS spectra of other hot-dust hosting
               stars. ID8 ($\tau$$\sim$2\%) is a $\sim$35 Myr old early-G type 
               star \citep{gorlova07}, P1121 ($\tau$$\sim$1\%) is a 
               $\sim$100 Myr old late-F type star \citep{gorlova04}, 
               BD+20 307 ($\tau$$\sim$4\%) is a $>$1 Gyr
               old spectroscopic binary composed of two 
               early-G type stars \citep{zuckerman08b}, 
               HD 113766A ($\tau$$\sim$1.5\%) is a $\sim$16 Myr old
               early-F type star \citep{chen06}, and EF Cha ($\tau$$\sim$0.1\%)
               is a $\sim$10 Myr old
               late-A type star \citep{rhee07b}. HD 113766A's IRS spectrum has been
               analyzed in great detail and is dominated by
               amorphous olivine and pyroxene features and crystalline olivine features
               \citep{chen06,lisse08}.}
\end{figure}

Excess infrared emission is detected towards HD 15407A by IRAS in the
12 and 25 $\mu$m bands \citep{oudmaijer92}. Archival Spitzer IRS spectroscopy of HD 15407A
(AOR 26122496, PI Fujiwara) confirms the excess (see Figure \ref{figirex}). We note
the presence of solid state dust emission features in the IRS spectrum. 
Figure \ref{figdusty} compares the IRS spectrum of HD 15407A with those
of other stars that host substantial quantities of terrestrial planet zone dust. 
We defer quantitative analysis of these spectral features to later
publications.

To estimate the dust temperature and the fraction of the stellar luminosity 
reradiated by the dust ($\tau$=L$_{\rm IR}$/L$_{\rm bol}$) we fit optical and 
near-infrared 
measurements out to K$_{\rm s}$-band with a synthetic stellar atmosphere spectrum 
\citep{hau99} and a blackbody to model the dust excess (see Figure \ref{figirex}). 
In this way we find that the fraction of stellar light reradiated by the circumstellar dust
is $\sim$5.7$\times$10$^{-3}$ and that the grains radiate with a blackbody temperature
of $\sim$500 K. Such grains will reside at a distance of $\sim$0.8 AU from HD 15407A.

We note that both the IRS spectrum and an IRAS upper limit at 60 $\mu$m
of 347 mJy (4$\sigma$) suggest that there is no cold dust orbiting
HD 15407A.





\subsection{Age of the HD 15407 System}
\label{secage}

\begin{figure}
 \begin{minipage}[t!]{91mm}
  \vskip -2.44in
  \includegraphics[width=91mm]{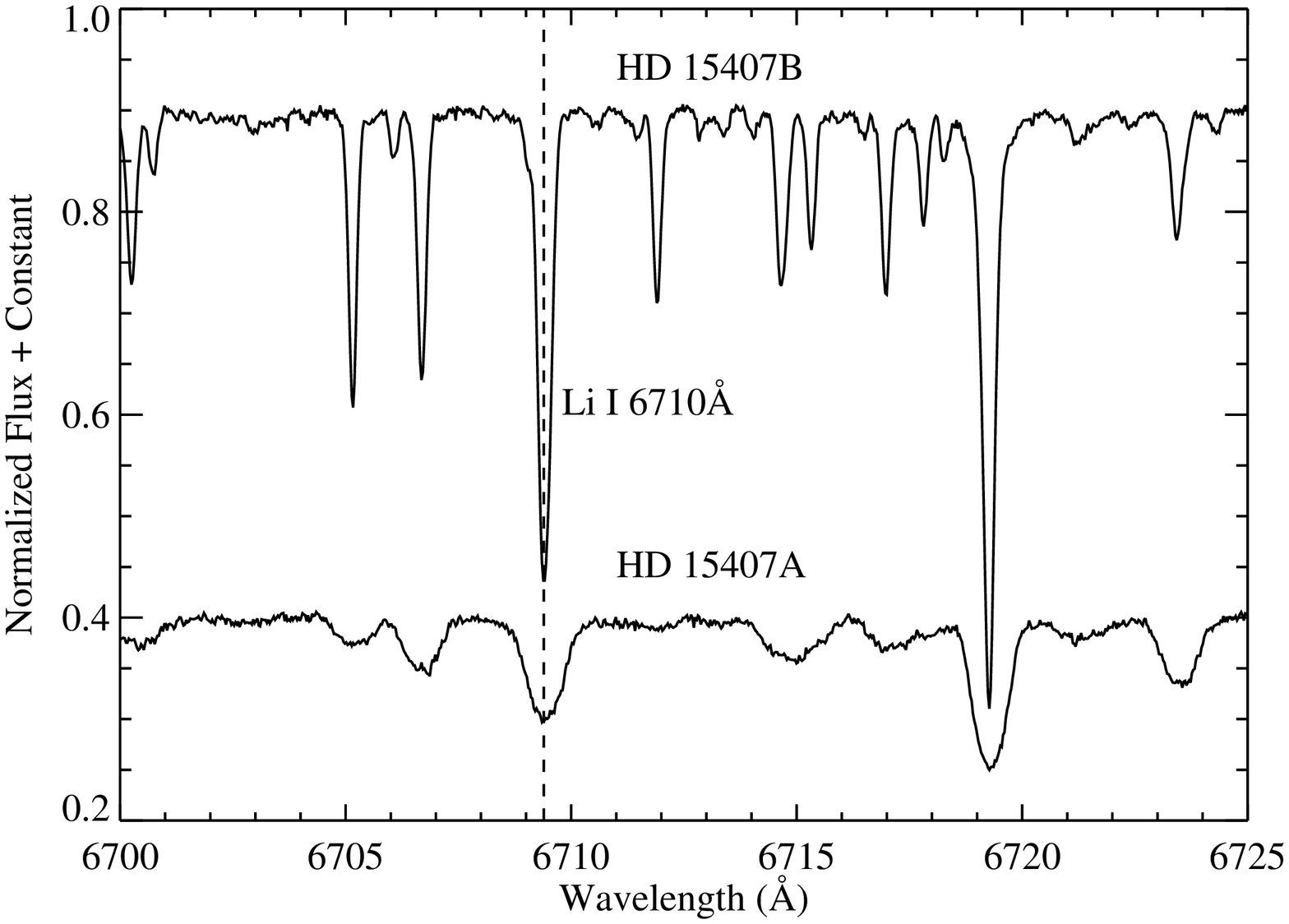}
 \end{minipage}
 \begin{minipage}[t!]{95mm}
  \includegraphics[width=95mm]{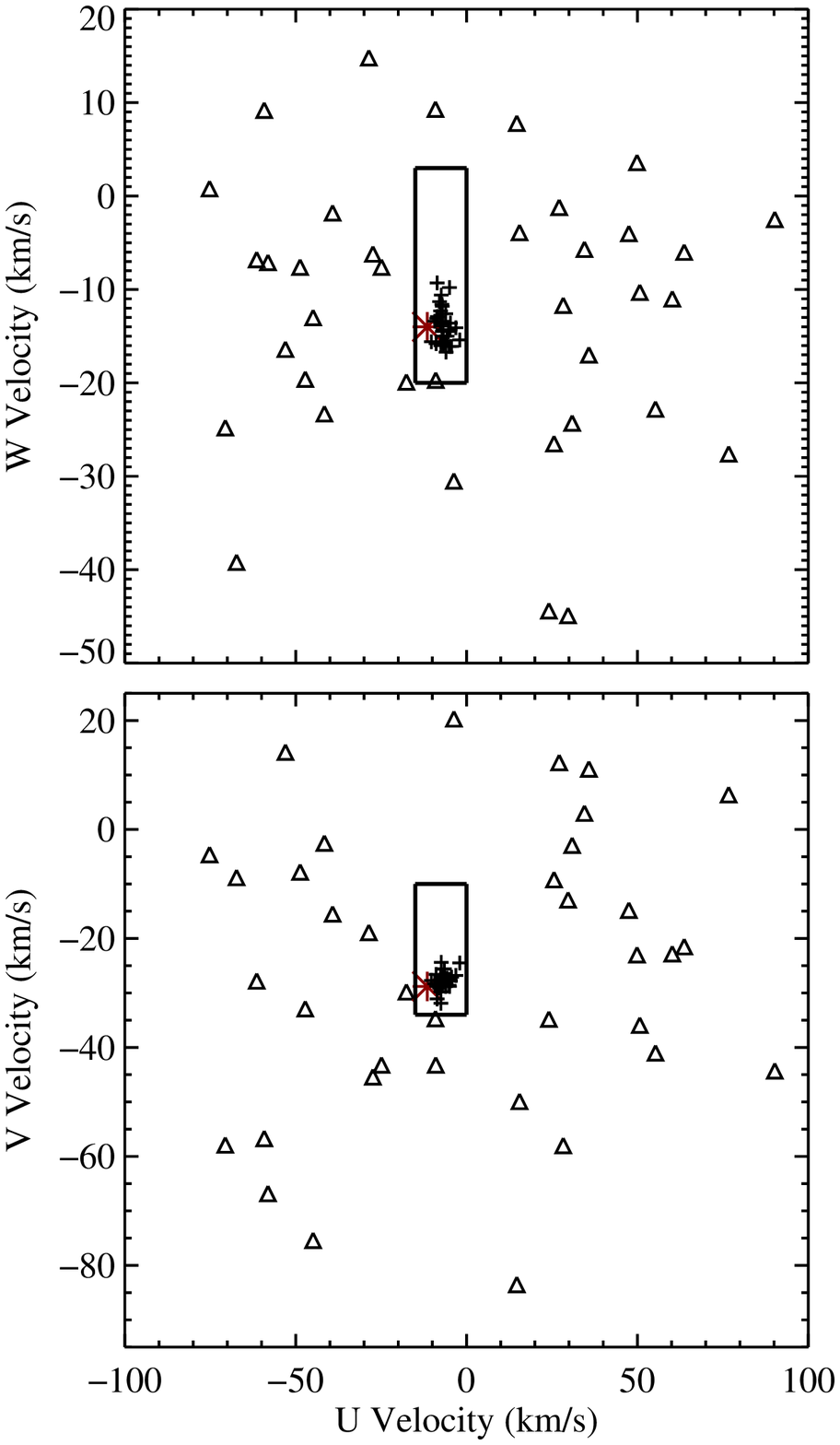}
 \end{minipage}
 \\
 \vskip -2.51in
 \hskip 0.1in
 \begin{minipage}[b!]{85mm}
  \includegraphics[width=85mm]{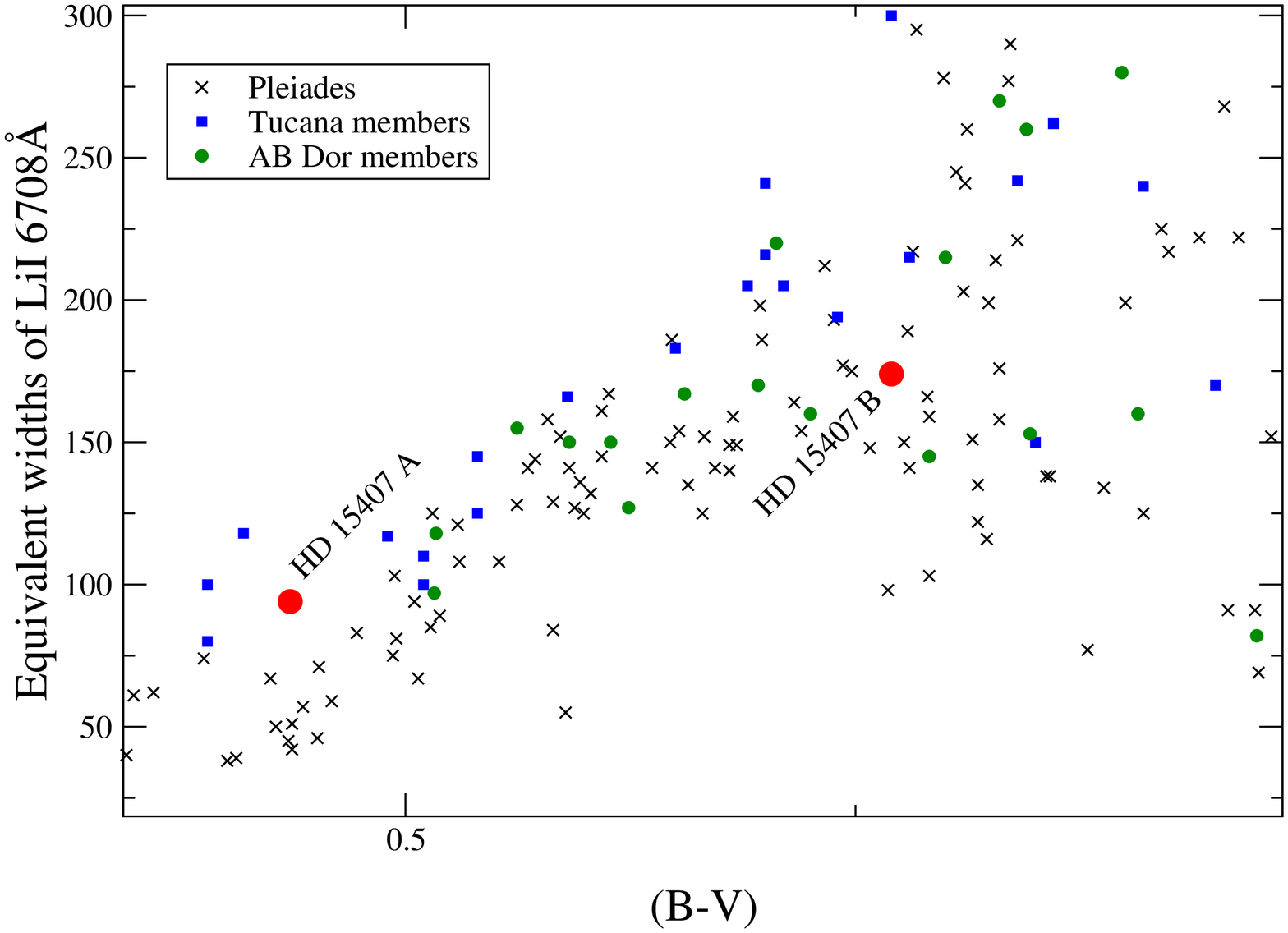}
 \end{minipage}
\caption{\label{figlith} {\it Top Left Panel:} Keck HIRES spectra obtained on UT 29 Nov 2009 and covering the Li~I $\lambda$6710
               (vacuum) region for HD 15407A and B. Strong lithium lines are
               detected in both stars (see Table \ref{tabpars}). These spectra are
               continuum normalized and offset
               by arbitrary values. Wavelengths in this figure are plotted in
               the heliocentric reference frame and are in vacuum.
               {\it Bottom Left Panel:} Comparison of the measured Li~I EWs of HD 15407A and B
               (Table \ref{tabpars}) to those of moving groups with similar age 
               \citep[adapted from Figure 3 of][]{zuckerman04}. The Pleiades age is $\sim$100
               Myr, the Tucana association age is $\sim$30 Myr, and the AB Dor association
               age is $\sim$80 Myr \citep[see text and][]{zuckerman04}.
               {\it Right Panels:} Comparison of the Galactic UVW space motion of the 
               HD 15407 system (red star symbol) with
               the UVW space motions of (1) the cluster of plus symbols
               near the center of the figures that 
               are AB Dor moving group members from \citet{zsb04} and (2) triangles that
               are white dwarfs with mean total ages comparable to that of the Sun 
               \citep{zuckerman03,bergeron01}.
                The rectangles
                represent the ``good'' UVW box for young stars as defined in 
                \citet{zuckerman04}.}
\end{figure}

To determine the age of the binary system we obtained and 
analyzed high-resolution
echelle spectra for both stars and searched for X-ray detections in the
literature. From these data we estimate the system
age from the lithium content in the stellar photospheres, velocity widths of 
absorption lines, Galactic space motion, and
chromospheric activity; details can be 
found in \citet{zuckerman04}. The lithium 6710 \AA\ absorption feature 
is strongly detected in the spectra of both
stars (Table \ref{tabpars} and Figure \ref{figlith}). 
Lithium content in a stellar atmosphere is mainly 
determined by the star's age and mass. These strong 
detections are consistent with 
roughly Solar mass stars of $\sim$30-100 Myr age (see Figure \ref{figlith}). 
We note that the lithium equivalent
width (EW) measured for HD 15407A ($\sim$94 m\AA ) is significantly 
stronger than the range of Li~I EWs  ($\sim$40-70 m\AA) 
quoted for similar spectral type Pleiads in Figure 3; this 
suggests that 
the HD 15407 system may be younger than Pleiades cluster stars.

\begin{figure}
 \centering
 \includegraphics[width=130mm]{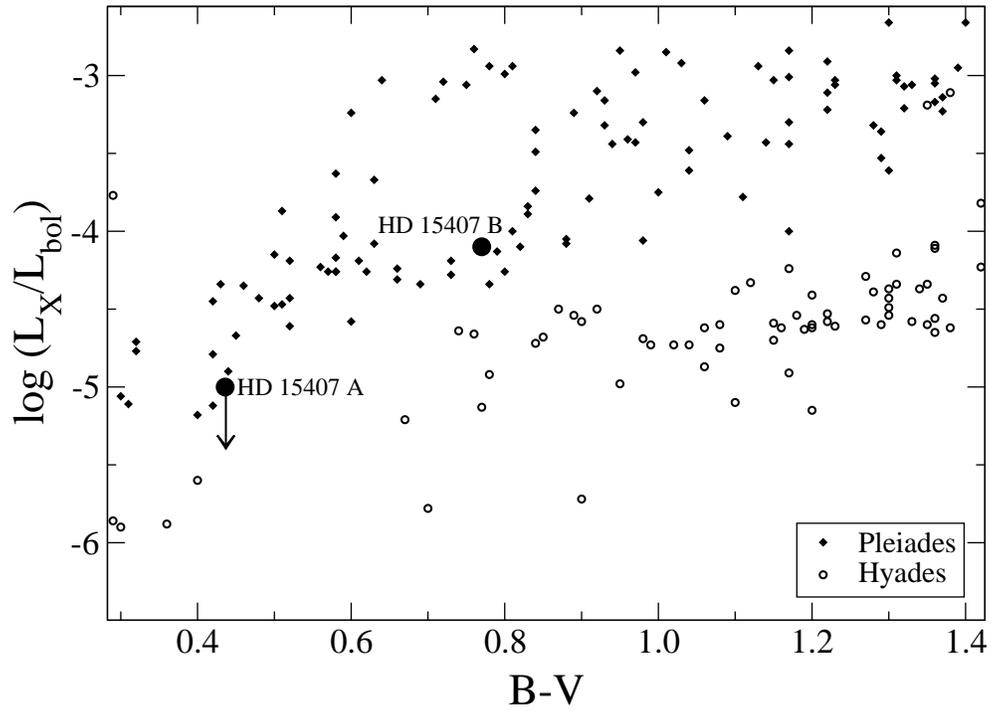}
\caption{\label{figxray} Comparison of the calculated fractional X-ray luminosities
               of HD 15407A and B (Table \ref{tabpars}) to those of two clusters of known age
               \citep[adapted from Figure 4 of][]{zuckerman04}. HD 15407A was not
               detected by the ROSAT satellite, the plotted value is an estimated upper
               limit (see text). The Pleiades age is $\sim$100 Myr while the Hyades
               age is $\sim$600 Myr.}
\end{figure}

The ROSAT All-Sky Survey (RASS) detected X-ray emission from HD 15407B. Analysis
of this X-ray emission (Table \ref{tabpars} and Figure \ref{figxray}) suggests a star of 
$\sim$100 Myr age \citep{zuckerman04}. We note that \citet{suchkov03} claim
a detection of HD 15407A by the RASS. Upon further
investigation, it is revealed that the putative detection has an angular
separation of $>$5$^{\prime}$
from HD 15407A, well outside the RASS positional uncertainties. We conclude that 
HD 15407A was not
detected by ROSAT. We estimate an upper limit for its fractional X-ray
luminosity in the following manner: HD 15407B is detected
by ROSAT with $\sim$4$\sigma$ significance. We take this level of X-ray
flux as our upper limit for HD 15407A. Thus, we derive the upper limit to
HD 15407A's fractional X-ray luminosity by scaling HD 15407B's
fractional X-ray luminosity by the ratio of the two star's total luminosities.
This upper limit is reported in Table \ref{tabpars} and plotted in
Figure \ref{figxray} where it is shown to be consistent with stars of
$\sim$100 Myr age.

Velocities of the HD 15407 system toward the center of our Milky Way galaxy,
around the Galactic Center, and perpendicular to the Galactic plane (U, V, W)
are calculated from Table \ref{tabpars} sky positions and proper motions, 
the Hipparcos measured distance
\citep[][where we assume HD 15407A and B are bound as is suggested by
their common proper motions 
and the optical echelle measured radial velocities, see Table \ref{tabpars}]{vanleeuwen07},
and heliocentric radial velocities.
UVW is computed for each component of the HD 15407 system
and these two results are averaged together to produce the values
reported in Table \ref{tabpars}.
Comparison of these computed UVW space motions to those of known
nearby stellar associations 
\citep[see Figure \ref{figlith} and e.g.,][and references therein]{zuckerman04,torres08}
suggests that the HD 15407 system is a member of the AB Dor moving
group (UVW$\sim$$-$8, $-$27, $-$14 km s$^{-1}$). The distance to the HD 15407
system and its position in the plane of the sky are consistent with it
being a member of the AB Dor moving group 
\citep[e.g., see Table 5 of][]{zuckerman04}. Age estimates for the AB Dor
moving group range from $\sim$50-120 Myr 
\citep[][and references therein]{zsb04,luhman05,torres08}.

Based on the above analysis, we suggest a best estimate age for the 
HD 15407 system of $\sim$80 Myr with a conservative range of plausible
ages from 60-120 Myr.
We note that the measured $v$sin$i$ for 
HD 15407B\footnote{$v$sin$i$ for both stars was measured from the
full-width at half-maximum depth (FHWM) of single absorption lines in the
Keck HIRES spectra; the intrinsic FWHM resolution for the HIRES setup used
was $\sim$6 \kms , a value we subtract in quadrature from the FWHM measured 
in the spectra when determining $v$sin$i$ \citep[see Equation 6 in][]{strassmeier90c}.},
$\sim$4 \kms , is surprisingly low
for a star of such age and spectral type \citep[][and references therein]{zuckerman04};
it is possible that we view HD 15407B in a pole-on geometry.




\section{Discussion}

Here we use our age estimate for the HD 15407 system to help
constrain the era of rocky planet formation in the terrestrial planet zone
around stars of Solar mass. We consider stars with fractional infrared luminosities 
$\tau$$\gtrsim$0.01 as
very dusty and likely to have undergone a recent collision between two planetary
embryos or planets such as that postulated to have formed Earth's Moon.
 \citet{rhee08} considered this problem, but
included the very dusty G-type star BD+20 307 in their analysis.
Subsequent to their paper, it was discovered that BD+20 307 is an old main
sequence star and hence its circumstellar dust is not related to terrestrial
planet formation \citep{weinberger08,zuckerman08b}.
\citet{rhee07b} and C. Melis {\it et al.}\ (2010, in preparation) have
investigated ``intermediate mass'' stars (spectral types late-B through early-F 
and masses from $\sim$8 to $\sim$1.5 M$_{\odot}$) 
with large masses of warm dust in their terrestrial planet zones.
The half dozen or so very dusty intermediate mass stars currently known have ages
between 10 and 30 Myr. This suggests that for such stars the era of most
intense accretion of material onto rocky planets is over by about 30 Myr.
Such a conclusion is similar to that drawn by \citet{currie08a} who
considered an ensemble of young intermediate mass stars but with only modest
amounts of warm dust.  We now show that the corresponding era for solar-mass stars, by 
contrast, lasts for $\sim$100 Myr.

Based on IRAS and Spitzer surveys we are aware of four adolescent 
Solar-type stars with $\tau$$\gtrsim$0.01:  
AB Dor member HD 15407A (this paper);
Pleiades member HD 23514 \citep{rhee08}; 
M47 member P1121 \citep{gorlova04}; 
and NGC 2547 member ID8 \citep{gorlova07}. 
The ages of these stars range between 35 Myr (NGC 2547) and about 
100 Myr (the other 3 stars). Interestingly, two of the four young, very dusty Sun-like stars
mentioned above are in wide binary systems (HD 23514, D. Rodriguez {\it et al.}\ 2010
in preparation; HD 15407, this work). The other two systems (ID8 and P1121)
have not been exhaustively searched for binary companions; confirmation
of any such binarity would be suggestive (although still in the realm
of statistics of small numbers) of a link between binarity and
catastrophic terrestrial planet collisions.

We note the relative absence of similar numbers of warm
dust stars with ages from $\sim$200 to 800 Myr. No stars in four stellar
associations with ages in this range $-$ the Hyades \citep{cieza08},
Praesepe \citep{gaspar09}, Carina-Near \citep{zuckerman06},
and Ursa Major \citep{king03} $-$ have substantial quantities of
warm dust in their terrestrial planet zone. Together these four groups
contain $\sim$200 Solar-type stars. In the field, IRAS has surveyed
many more stars with ages of $\sim$200 to 800 Myr compared
to the 30 to 100 Myr range, yet none in the former age range
have been identified with luminous warm dust.

We estimate the frequency of catastrophic accretion events during
rocky planet formation in the terrestrial
planet zone by considering the detection rate of very dusty stars
in the field and in young clusters. Very dusty nearby field stars were better
investigated in the all-sky IRAS survey than with Spitzer's pointed
capabilities. As noted by \citet{rhee08}, IRAS could have detected a main
sequence dwarf with $\tau$$\gtrsim$0.01 
and spectral type between F4 and K0 out to about 150 pc.  
There are 400 or so such field stars with ages $\sim$100 Myr that IRAS
could have detected, but of these, only HD 15407A has 
sufficient dust to actually have been detected by IRAS.  
Similarly, among the young clusters studied with Spitzer $-$
NGC 2232 \citep{currie08a}, M47 \citep{gorlova04}, 
NGC 2547 \citep{gorlova07}, IC 2391 \citep{siegler07}, 
and the Pleiades \citep{stauffer05,gorlova06} $-$
the three very dusty cluster stars listed previously
represent about one Solar-type star in 200
\citep[roughly consistent with the
determination for Solar-type cluster stars by][]{balog09}. 

Thus, the high $\tau$
warm dust phenomenon manifests itself at about one
adolescent Solar-type star in 300. 
If all F4-K0 stars display this phenomenon
as adolescents, then the lifetime of the phenomenon at a typical Sun-like
star is about 300,000 years.

To understand this frequency of the warm dust phenomenon we consider a dust
production model similar to that described by \citet{rhee08} for HD 23514.
\citet{rhee08} analyzed existing observations through October 2007 with the
aid of a model of colliding planetary embryos due to C. Agnor and E. Asphaug 
\citep{agnor99,agnor04,asphaug06}
and earlier researchers. The idea is that following a catastrophic collision
of two rocky planetary embryos or planets, fragmented debris covering a range
of sizes goes into orbit around a central star. \citet{rhee08} assumed that
collisions between debris would be frequent enough to establish
an equilibrium size distribution where the number of particles of
radius ``$a$'' is proportional to $a^{-3.5}$.
In this situation, both the mass
carried by particles of radius $a$ and the collision time are proportional to
the square root of $a$. Most of the mass is carried by large fragments while
small particles are responsible for most of the excess infrared luminosity and
collide the fastest. The radius of the smallest
dust particles in orbit around a very dusty star is set by radiative blowout and,
for HD 15407A, will
be $\sim$1 $\mu$m.  The collision time for such particles will be about 1 year
divided by 10$\times$L$_{\rm IR}$/L$_{\rm bol}$ 
or $\sim$10 years. If the largest initial fragments
of a catastrophic collision have radii $\sim$100 m, then their collisional
lifetime is $\sim$100,000 years which is within a factor of 3 of the
lifetime (estimated in the previous paragraph) required if all Solar-type
stars are forming rocky planets in their terrestrial planet zone. Should the
formation of rocky terrestrial planets involve more
than a single catastrophic collision, then the model-estimated warm dust
phenomenon lifetime could be more than 100,000 years
at each Sun-like star. As noted by 
\citet{rhee08}, the total mass lost over this interval of time $-$ due to a
collisional cascade followed by radiative blowout $-$ will be about the mass
of Earth's Moon.

\section{Conclusions}

Based on Keck HIRES optical spectroscopy, X-ray flux, and Galactic UVW space motions 
we estimate the age
of the HD 15407 binary system to be $\sim$80 Myr and identify the system as likely
belonging to the AB Dor moving group. Collecting the full sample
of known very dusty debris-disk hosting stars, we find
that the era of final catastrophic mass accretion for rocky planets 
around stars of roughly Solar mass occurs later than around intermediate mass stars.
Provided that rocky planet formation in the terrestrial planet zone of
most Solar mass stars is accompanied by at least one and possibly a few
catastrophic collisions of planet-size objects within a 100 Myr interval, then
the detection rate and age range of very dusty Solar-type stars $-$ observed 
to be about one in 300 for ages between 30 and 100 Myr $-$ can be accounted 
for. Our conclusion is that existing observations of
such dusty stars indicates that most Sun-like stars are accompanied by rocky
planets in the terrestrial planet zone and that collisions among such forming
planets will occur, as is expected in theoretical simulations of planetary
origins.

\acknowledgments

We are indebted to George Herbig for obtaining the HIRES data for us.
We thank Eric Becklin for pointing out this stellar system to us.
Some of the data presented herein were obtained at the 
W.M. Keck Observatory, which is operated as a scientific partnership among the 
California Institute of Technology, the University of California and the 
National Aeronautics and Space Administration. The Observatory was made possible
by the generous financial support of the W.M. Keck Foundation. 
This research has made use of the VizieR service and of data products from
the Two Micron All Sky Survey. This research was supported in part by a NASA
grant to UCLA.



{\it Facilities:} \facility{IRAS ()}, \facility{Keck:I (HIRES)}, \facility{Spitzer (IRS)}

\clearpage





\clearpage

\clearpage

\clearpage

\begin{table}
\caption{Parameters of the HD 15407 binary star system \label{tabpars}}
\centering
\begin{tabular}{ccc} 
\tableline
\multicolumn{3}{c}{Individual Stars} \\
\tableline
\tableline
 & HD 15407A & HD 15407B \\
\tableline
RA (J2000) & 02 30 50.65 & 02 30 48.58 \\
DEC (J2000) & +55 32 54.4 & +55 33 06.4 \\
Sp. Type & F5~V & K2~V \\
Vmag & 6.95 & 9.58 \\
T$_{\rm eff}$ (K) & 6500$\pm$100 & 4830$\pm$150 \\
pmRA (mas yr$^{-1}$) & +80.6 $\pm$ 1.0 & +83.1 $\pm$ 1.0 \\
pmDE (mas yr$^{-1}$) & $-$96.7 $\pm$ 1.1 & $-$93.6 $\pm$ 1.1 \\
RV (\kms ) & $-$10.44 $\pm$ 0.39 & $-$10.02 $\pm$ 0.17 \\
Li~I $\lambda$6710 EW (m\AA) & 94$\pm$4 & 174$\pm$5 \\
$v$sin$i$ (\kms ) & 20 & 4 \\
L$_{\rm x}$/L$_{\rm bol}$ & $<$10$^{-5.0}$ & 10$^{-4.1}$ \\
\tableline
\multicolumn{3}{c}{System} \\
\tableline
\tableline
Distance from Earth (pc) & \multicolumn{2}{c}{54.7$\pm$2.1} \\
Separation Between Stars & \multicolumn{2}{c}{21.25$\arcsec$ (projected separation of 1160 AU)} \\
UVW Space Motions (\kms ) & \multicolumn{2}{c}{$-$11.5, $-$28.8, $-$14.0} \\
Age (Myr) & \multicolumn{2}{c}{80$\mathop{}_{-20}^{+40}$} \\
\tableline
\end{tabular}
\tablecomments{Epoch J2000 positions are from the 2MASS catalog \citep{cutri03}. 
The highest fidelity proper motions come from the Tycho-2 catalog \citep{hog00} 
for HD 15407A and from the UCAC3 catalog \citep{zacharias10} for HD 15407B. 
Radial velocities (RV) for both stars were measured by cross-correlating our HIRES spectra
with HIRES spectra of a star from Table 2 of \citet{nidever02}. See the text for discussion 
of other stellar parameters. }

\end{table}







\begin{thebibliography}{36}
\expandafter\ifx\csname natexlab\endcsname\relax\def\natexlab#1{#1}\fi

\bibitem[\protect\astroncite{{Agnor} \& {Asphaug}}{2004}]{agnor04}
{Agnor}, C. \& {Asphaug}, E. 2004, {\em \apjl\/}, {\bf 613}, L157

\bibitem[\protect\astroncite{{Agnor} {\em et~al.\/}}{1999}]{agnor99}
{Agnor}, C.~B., {Canup}, R.~M., \& {Levison}, H.~F. 1999, {\em Icarus\/}, {\bf
  142}, 219

\bibitem[\protect\astroncite{{Asphaug} {\em et~al.\/}}{2006}]{asphaug06}
{Asphaug}, E., {Agnor}, C.~B., \& {Williams}, Q. 2006, {\em \nat\/}, {\bf 439},
  155

\bibitem[\protect\astroncite{{Balog} {\em et~al.\/}}{2009}]{balog09}
{Balog}, Z., {\em et~al.\/} 2009, {\em \apj\/}, {\bf 698}, 1989

\bibitem[\protect\astroncite{{Bergeron} {\em et~al.\/}}{2001}]{bergeron01}
{Bergeron}, P., {Leggett}, S.~K., \& {Ruiz}, M.~T. 2001, {\em \apjs\/}, {\bf
  133}, 413

\bibitem[\protect\astroncite{{Chen} {\em et~al.\/}}{2006}]{chen06}
{Chen}, C.~H., {\em et~al.\/} 2006, {\em \apjs\/}, {\bf 166}, 351

\bibitem[\protect\astroncite{{Cieza} {\em et~al.\/}}{2008}]{cieza08}
{Cieza}, L.~A., {Cochran}, W.~D., \& {Augereau}, J. 2008, {\em \apj\/}, {\bf
  679}, 720

\bibitem[\protect\astroncite{{Currie} {\em et~al.\/}}{2008}]{currie08a}
{Currie}, T., {Kenyon}, S.~J., {Balog}, Z., {Rieke}, G., {Bragg}, A., \&
  {Bromley}, B. 2008, {\em \apj\/}, {\bf 672}, 558

\bibitem[\protect\astroncite{{Cutri} {\em et~al.\/}}{2003}]{cutri03}
{Cutri}, R.~M., {\em et~al.\/} 2003, {\em {2MASS All Sky Catalog of point
  sources.}\/}, The IRSA 2MASS All-Sky Point Source Catalog, NASA/IPAC Infrared
  Science Archive.~http://irsa.ipac.caltech.edu/applications/Gator/

\bibitem[\protect\astroncite{{G{\'a}sp{\'a}r} {\em et~al.\/}}{2009}]{gaspar09}
{G{\'a}sp{\'a}r}, A., {\em et~al.\/} 2009, {\em \apj\/}, {\bf 697}, 1578

\bibitem[\protect\astroncite{{Gorlova} {\em et~al.\/}}{2007}]{gorlova07}
{Gorlova}, N., {Balog}, Z., {Rieke}, G.~H., {Muzerolle}, J., {Su}, K.~Y.~L.,
  {Ivanov}, V.~D., \& {Young}, E.~T. 2007, {\em \apj\/}, {\bf 670}, 516

\bibitem[\protect\astroncite{{Gorlova} {\em et~al.\/}}{2006}]{gorlova06}
{Gorlova}, N., {Rieke}, G.~H., {Muzerolle}, J., {Stauffer}, J.~R., {Siegler},
  N., {Young}, E.~T., \& {Stansberry}, J.~H. 2006, {\em \apj\/}, {\bf 649},
  1028

\bibitem[\protect\astroncite{{Gorlova} {\em et~al.\/}}{2004}]{gorlova04}
{Gorlova}, N., {\em et~al.\/} 2004, {\em \apjs\/}, {\bf 154}, 448

\bibitem[\protect\astroncite{{Hauschildt} {\em et~al.\/}}{1999}]{hau99}
{Hauschildt}, P.~H., {Allard}, F., \& {Baron}, E. 1999, {\em \apj\/}, {\bf
  512}, 377

\bibitem[\protect\astroncite{{H{\o}g} {\em et~al.\/}}{2000}]{hog00}
{H{\o}g}, E., {\em et~al.\/} 2000, {\em \aap\/}, {\bf 355}, L27

\bibitem[\protect\astroncite{{King} {\em et~al.\/}}{2003}]{king03}
{King}, J.~R., {Villarreal}, A.~R., {Soderblom}, D.~R., {Gulliver}, A.~F., \&
  {Adelman}, S.~J. 2003, {\em \aj\/}, {\bf 125}, 1980

\bibitem[\protect\astroncite{{Lisse} {\em et~al.\/}}{2008}]{lisse08}
{Lisse}, C.~M., {Chen}, C.~H., {Wyatt}, M.~C., \& {Morlok}, A. 2008, {\em
  \apj\/}, {\bf 673}, 1106

\bibitem[\protect\astroncite{{Lisse} {\em et~al.\/}}{2009}]{lisse09}
{Lisse}, C.~M., {Chen}, C.~H., {Wyatt}, M.~C., {Morlok}, A., {Song}, I.,
  {Bryden}, G., \& {Sheehan}, P. 2009, {\em \apj\/}, {\bf 701}, 2019

\bibitem[\protect\astroncite{{Luhman} {\em et~al.\/}}{2005}]{luhman05}
{Luhman}, K.~L., {Stauffer}, J.~R., \& {Mamajek}, E.~E. 2005, {\em \apjl\/},
  {\bf 628}, L69

\bibitem[\protect\astroncite{{Melis} {\em et~al.\/}}{2009}]{melis09}
{Melis}, C., {Zuckerman}, B., {Song}, I., {Rhee}, J.~H., \& {Metchev}, S. 2009,
  {\em \apj\/}, {\bf 696}, 1964

\bibitem[\protect\astroncite{{Nidever} {\em et~al.\/}}{2002}]{nidever02}
{Nidever}, D.~L., {Marcy}, G.~W., {Butler}, R.~P., {Fischer}, D.~A., \& {Vogt},
  S.~S. 2002, {\em \apjs\/}, {\bf 141}, 503

\bibitem[\protect\astroncite{{Oudmaijer} {\em et~al.\/}}{1992}]{oudmaijer92}
{Oudmaijer}, R.~D., {van der Veen}, W.~E.~C.~J., {Waters}, L.~B.~F.~M.,
  {Trams}, N.~R., {Waelkens}, C., \& {Engelsman}, E. 1992, {\em \aaps\/}, {\bf
  96}, 625

\bibitem[\protect\astroncite{{Rhee} {\em et~al.\/}}{2007}]{rhee07b}
{Rhee}, J.~H., {Song}, I., \& {Zuckerman}, B. 2007, {\em \apj\/}, {\bf 671},
  616

\bibitem[\protect\astroncite{{Rhee} {\em et~al.\/}}{2008}]{rhee08}
--- 2008, {\em \apj\/}, {\bf 675}, 777

\bibitem[\protect\astroncite{{Siegler} {\em et~al.\/}}{2007}]{siegler07}
{Siegler}, N., {\em et~al.\/} 2007, {\em \apj\/}, {\bf 654}, 580

\bibitem[\protect\astroncite{{Stauffer} {\em et~al.\/}}{2005}]{stauffer05}
{Stauffer}, J.~R., {\em et~al.\/} 2005, {\em \aj\/}, {\bf 130}, 1834

\bibitem[\protect\astroncite{{Strassmeier} {\em
  et~al.\/}}{1990}]{strassmeier90c}
{Strassmeier}, K.~G., {Fekel}, F.~C., {Bopp}, B.~W., {Dempsey}, R.~C., \&
  {Henry}, G.~W. 1990, {\em \apjs\/}, {\bf 72}, 191

\bibitem[\protect\astroncite{{Suchkov} {\em et~al.\/}}{2003}]{suchkov03}
{Suchkov}, A.~A., {Makarov}, V.~V., \& {Voges}, W. 2003, {\em \apj\/}, {\bf
  595}, 1206

\bibitem[\protect\astroncite{{Torres} {\em et~al.\/}}{2008}]{torres08}
{Torres}, C.~A.~O., {Quast}, G.~R., {Melo}, C.~H.~F., \& {Sterzik}, M.~F. 2008, {\em
  {Young Nearby Loose Associations}\/},  757--+

\bibitem[\protect\astroncite{{van Leeuwen}}{2007}]{vanleeuwen07}
{van Leeuwen}, F. 2007, {\em \aap\/}, {\bf 474}, 653

\bibitem[\protect\astroncite{{Vogt} {\em et~al.\/}}{1994}]{vogt94}
{Vogt}, S.~S., {\em et~al.\/} 1994, in {\em Proc. SPIE Instrumentation in
  Astronomy VIII, David L. Crawford; Eric R. Craine; Eds., Volume 2198, p.
  362\/}, edited by D.~L. {Crawford} \& E.~R. {Craine}, vol. 2198 of {\em
  Presented at the Society of Photo-Optical Instrumentation Engineers (SPIE)
  Conference\/},  362

\bibitem[\protect\astroncite{{Weinberger}}{2008}]{weinberger08}
{Weinberger}, A.~J. 2008, {\em \apjl\/}, {\bf 679}, L41

\bibitem[\protect\astroncite{{Zacharias} {\em et~al.\/}}{2010}]{zacharias10}
{Zacharias}, N., {\em et~al.\/} 2010, {\em \aj\/}, {\bf 139}, 2184

\bibitem[\protect\astroncite{{Zuckerman} {\em et~al.\/}}{2006}]{zuckerman06}
{Zuckerman}, B., {Bessell}, M.~S., {Song}, I., \& {Kim}, S. 2006, {\em
  \apjl\/}, {\bf 649}, L115

\bibitem[\protect\astroncite{{Zuckerman} {\em et~al.\/}}{2008}]{zuckerman08b}
{Zuckerman}, B., {Fekel}, F.~C., {Williamson}, M.~H., {Henry}, G.~W., \&
  {Muno}, M.~P. 2008, {\em \apj\/}, {\bf 688}, 1345

\bibitem[\protect\astroncite{{Zuckerman} {\em et~al.\/}}{2003}]{zuckerman03}
{Zuckerman}, B., {Koester}, D., {Reid}, I.~N., \& {H{\"u}nsch}, M. 2003, {\em
  \apj\/}, {\bf 596}, 477

\bibitem[\protect\astroncite{{Zuckerman} \& {Song}}{2004}]{zuckerman04}
{Zuckerman}, B. \& {Song}, I. 2004, {\em \araa\/}, {\bf 42}, 685

\bibitem[\protect\astroncite{{Zuckerman} {\em et~al.\/}}{2004}]{zsb04}
{Zuckerman}, B., {Song}, I., \& {Bessell}, M.~S. 2004, {\em \apjl\/}, {\bf
  613}, L65

\end{thebibliography}
\end{document}